\documentclass[runningheads]{llncs}
\usepackage[T1]{fontenc}
\usepackage{csquotes}
\usepackage{float}
\usepackage{graphicx}
\usepackage{caption}
\usepackage{subcaption}
\usepackage{tabularx}
\usepackage{booktabs}
\usepackage{subfig}

\usepackage{enumitem}

\newenvironment{tightitemize}{
\begin{itemize}[leftmargin=0pt]
  \setlength{\itemsep}{0pt}
  \setlength{\parskip}{-6pt}
  \setlength{\parsep}{0pt}
}{\end{itemize}
}

\begin{document}
\title{From Guidelines to Governance: A Study of AI Policies in Education}
%
%
\author{Aashish Ghimire \and
John Edwards }
%
%
\institute{Utah State University, \\
Old Main Hill \\
Logan , UT, USA \\
}
\maketitle              
\begin{abstract}
Emerging technologies like generative AI tools, including ChatGPT, are increasingly utilized in educational settings, offering innovative approaches to learning while simultaneously posing new challenges. This study employs a survey methodology to examine the policy landscape concerning these technologies, drawing insights from 102 high school principals and higher education provosts. Our results reveal a prominent policy gap: the majority of institutions lack specialized guidelines for the ethical deployment of AI tools such as ChatGPT. Moreover, we observed that high schools are less inclined to work on policies
than higher educational institutions. Where such policies do exist, they often overlook crucial issues, including student privacy and algorithmic transparency. Administrators overwhelmingly recognize the necessity of these policies, primarily to safeguard student safety and mitigate plagiarism risks. Our findings underscore the urgent need for flexible and iterative policy frameworks in educational contexts.

\end{abstract}
\section{Introduction}

With the rapid advancement of technology, generative artificial intelligence (AI) tools, particularly Large Language Models (LLMs) like ChatGPT, are increasingly being adopted in various sectors, including education. These technologies offer promising avenues for pedagogical innovation, personalized learning, and administrative efficiency. However, their integration into educational settings is not without challenges, particularly concerning ethical considerations. Issues related to student privacy, data security, algorithmic transparency, and accountability are growing areas of concern.

While the application of these tools offers numerous advantages, the absence of comprehensive policy frameworks governing their ethical use in education can lead to unintended negative consequences. Inadequate policies may expose students to risks such as data misuse, algorithmic bias, and academic dishonesty. Educational institutions, thus, find themselves at a crossroads, balancing the potential benefits of emerging technologies against ethical and legal ramifications.
Artificial Intelligence (AI) in education has garnered significant attention, leading to an increase in scholarly inquiries. The focus of these studies predominantly revolves around the implementation and efficacy of AI-powered educational tools, often sidelining essential discourses on policy, ethics, and administrative perspectives.

Given the escalating integration of AI tools like ChatGPT in educational settings, there is an imperative need to understand the current landscape of ethical policies, or the lack thereof, governing their use. Understanding administrators' attitudes and perceptions towards these ethical considerations is crucial for formulating effective policies that can guide responsible AI adoption in education.

The research focuses on addressing the following questions:

\begin{enumerate}
    \item[RQ1] What is the current landscape of policies related to Generative AI in educational settings and what do these policies cover? 
    \item[RQ2] What are the perceived needs for future policy formulation in relation to Generative AI, and what recommendations can be made for an effective ethical framework?

\end{enumerate}

To answer these questions, this study adopts a mixed-methods research design, incorporating both quantitative and qualitative data collected via a survey of over 100 educational administrators in the United States.

The remainder of this paper is organized as follows: Section 2 outlines the methodology, Section 3 presents the findings, Section 4 offers a discussion, and Section 5 concludes with recommendations for policy formulation.

\section{Related Work}

The integration of artificial intelligence (AI) in education is evolving rapidly, necessitating a multidimensional understanding of its applications, ethical considerations, governance frameworks, and pedagogical impacts. This section synthesizes key contributions across these areas, providing a coherent overview of the current research landscape.

\subsection{Applications and Trends in AI in Education}

Recent studies highlight significant advancements and trends in AI's educational applications. Zhai et al.\cite{Zhai2021Review} and Chen et al.\cite{chen2022two} have identified critical research areas, including the Internet of Things, swarm intelligence, deep learning, and the application of natural language processing and neural networks in education. Works by Pradana et al.\cite{pradana2023discussing}, Lo\cite{lo2023impact}, and Choi et al.\cite{choi2023chatgpt} emphasize the diverse applications of AI tools, notably ChatGPT, and the importance of addressing gaps in ethical and social considerations. Flogie and Krabonja\cite{flogie2023artificial} discuss the challenges and models for integrating AI into teaching, underscoring the field's evolving nature and the need for comprehensive research covering technological, ethical, and administrative aspects.

\subsection{Ethical Challenges and Frameworks}

The ethical implications of AI in education are complex, involving considerations of fairness, transparency, and privacy. Holmes et al.\cite{holmes2021ethics}, Akgun and Greenhow\cite{akgun2021artificial}, and Adams et al.\cite{Adams2023Ethical} discuss the ethical challenges in deploying AI in educational settings. Halaweh et al.\cite{halaweh2023chatgpt} and Sullivan et al.\cite{sullivan2023chatgpt} propose frameworks for responsible implementation, emphasizing the need for policies that ensure student safety and academic integrity. Chiu\cite{chiu2023impact} and Kooli~\cite{Kooli2023Chatbots} highlight the lack of policy considerations, calling for a balanced approach to leveraging AI's benefits while mitigating its risks.

\subsection{Accountability, Fairness, and Governance}

The governance of AI in education involves balancing technological benefits with ethical risks. Garshi et al.\cite{garshi2020smart}, Berendt et al.\cite{berendt2020ai}, and Filgueiras~\cite{filgueiras2023artificial} explore and propose frameworks for accountability and human rights in smart classrooms. Li and Gu~\cite{li2023risk} present a risk framework for Human-Centered AI, emphasizing accountability and bias. Memarian and Doleck~\cite{memarian2023fairness}, Nigam et al~\cite{nigam2021systematic},Sahlgren~\cite{sahlgren2023politics}, and Gillani et al.~\cite{Gillani2023Unpacking} discuss the challenges of fairness and transparency,  necessity of security and privacy, ethical concerns, advocating for human-centered and politically aware governance models. Uunona and Goosen focus on leveraging ethical values in AI-powered online learning applications, particularly in the Namibian educational context~\cite{Uunona2023Leveraging}.

\subsection{Policy Guidelines and Implications}

The development of AI-specific policy guidelines is critical for ethical integration into educational systems. Miao et al.\cite{miao2021ai} and Chan\cite{Chan2023Comprehensive} have contributed to guiding policymakers, though existing technology policies provided by CoSN and broader guidelines from organizations like IEEE and the European Commission ~\cite{cosnCoSNIssues2019,chatila2018ethically,ai2019high,cosnCoSNIssues2023} lack the granularity needed for AI and  fall short in addressing AI's unique challenges. This underscores the need for more detailed and AI-focused educational policies.

\subsection{Pedagogical Approaches and Curriculum Design}

Pedagogical innovation is essential for integrating AI into education effectively. Ali et al.\cite{ali2023constructing} advocate for AI literacy in curricula, while Sattelmaier and Pawlowski\cite{sattelmaier2023towards} propose a competence framework for incorporating generative AI into school curricula. Ouyang et al.~\cite{ouyang2021artificial} present a framework for understanding AI's role in learning, highlighting the shift towards learner-centric models.

\subsection{Multidisciplinary Perspectives}

A multidisciplinary approach is vital for understanding AI's impact on education. Dwivedi et al.\cite{dwivedi2023so} and Baidoo-Anu and Owusu Ansah\cite{baidoo2023education} combine insights from various fields, addressing the capabilities and challenges of AI. Whalen and Mouza~\cite{whalen2023chatgpt} emphasize the need for ethical uses.

\begin{figure}
    \centering
\fbox{
\begin{minipage}{0.99\columnwidth}
\textbf{Demography}

Four questions: years of experience in education administration, number of students, faculty size, school public or private.

\vspace{0.7em}
\textbf{Current landscape of policies (RQ1)}
        \begin{itemize}
        \item Policy on emerging technologies in place? [\textit{Have policy/Working on a policy/No policy and not working on one/Don't know}] \textbf{\textsuperscript{M}}
        \item How necessary is it to have a policy? [\textit{Not/Somewhat/Very necessary}]
        
        \hspace{0.3em}
        
        \textbf{The following questions are shown if they have an AI policy :}
        \item Current policies adequate? [\textit{Likert scale: Strongly disagree to Strongly agree}]
        \item Policy specifically mentions LLMs such as ChatGPT? [\textit{Yes/No/Unsure}]
        \item  Which of the following elements are covered in your policy? [\textit{Student privacy/Algorithmic transparency/Bias mitigation/Accountability mechanisms/Plagiarism/Other - Free entry} ]\textbf{\textsuperscript{M}}
        \item Primary motivations for implementing or revising policy governing use of these AI tools in the classroom? [\textit{Stopping/Plagiarism/Ensuring student safety/Compliance with regulations/Ethical considerations/Research integrity**/Parental demand/Teachers' demand/Other - Free entry}]\textbf{\textsuperscript{M}}
        \end{itemize}
\textbf{Perceived needs and recommendations (RQ2)}
        \begin{itemize}
        \item  Who should be primarily responsible for formulating policy?  [\textit{School administration/School board*/Teachers/Parent-Teacher Association*/Higher Education board**/Faculty Senate**/Independent body/Students/Other (free entry)}]\textbf{\textsuperscript{M}}
        \item How much autonomy should individual schools have in setting or implementing policies?  [\textit{None/Some/Moderate/Most/All}]
        \item  How much autonomy should individual teachers have in setting or implementing policies?  [\textit{None/Some/Moderate/Most/All}]
        \item In which areas should policies focus?  [\textit{Stopping Plagiarism/Ensuring student safety/Compliance with regulations/Pedagogical innovation/Research purposes**/Ethical considerations/Student engagement/Using these tools to help reduce the teacher's workload/Other (free entry)}] \textbf{\textsuperscript{M}}
        \item What kind of support or resources would be helpful for your institution to create and implement policies?  [\textit{Professional development/Consultation with tech companies/Consultation with legal or ethics experts/Funding or resources/Model policies or guidelines from successful schools or districts/Other (free entry)}] \textbf{\textsuperscript{M}}
        
        \item Are there any specific policy components that you believe should be included in guidelines?  [\textit{Free entry}]
        \end{itemize}
\textbf{Other Questions (RQ4)}
    \begin{itemize}
        \item Overall opinion of LLMs?  [\textit{Likert scale : Dislike a great deal to Like a great deal} ]
        \item Do you have a policy that allows for punishing students based on results from AI-detection tools?[
        \textit{Such tools are banned/Such tools are used to narrow down but not as only factor to decide/Student can be punished based on the result of such tool-detected AI content. [Tool name]}]
        \item Additional comments. [\textit{Free entry}]
        \item Interested in a follow-up interview?
        \hspace{1 em}
        \end{itemize}
    \textbf{Options: \textsuperscript{M} Multiple selection allowed; * only to high school administrators, ** only to higher ed administrators}
        \hspace{0.5em}
\end{minipage}
}
\caption{Categories and Related Survey Questions. Most questions are condensed due to space constraints.}
    \label{fig:categories_and_questions}
\end{figure}


    

\section{Methodology}

To gain insights into the current policy landscape regulating the use of AI tools such as ChatGPT in educational settings, as well as to understand the attitudes of educational administrators toward these policies, this study employed a survey. This survey, administered across a diverse array of educational institutions, consists of a mix of multiple-choice questions, Likert-scale questions, and free-form text entries. The survey was specifically designed to discover the current landscape of policies related to Generative AI in educational settings and the perceived needs for future policy formulation in relation to Generative AI. Influenced by prior research such as Nguyen et al.~\cite{Nguyen2023Ethical} and Adams et al.~\cite{Adams2023Ethical}, the survey covered commonly identified policy areas and offered respondents the opportunity to express additional concerns and policy suggestions through free-form text. Figure~\ref{fig:categories_and_questions} outlines the questions included in the survey. Some options and language of questions were slightly changed to tailor the survey to high school and higher education administrators.

The primary focus of this study was on two groups of educational administrators: high school principals and academic officers or provosts in higher education institutions. These individuals were selected based on their pivotal roles in policy formulation and implementation within their respective organizations. The study garnered responses from over 100 administrators.

\subsubsection{Data Collection Instrument}
The survey, structured to align with the four primary objectives of the study, was hosted on the Qualtrics platform. It featured both closed-ended questions, aimed at capturing quantifiable metrics, and open-ended questions designed to explore the subjective viewpoints and rationales of administrators.

For distribution, we utilized a publicly available directory to identify and reach out to high school principals. We downloaded the mailing list of school administrators from the state education board’s website. Conversely, for higher education institutions, we employed a manually curated mailing list. To do this, we first obtained a list of all higher education institutes in the states, went to their websites, and looked up their provost’s or chief academic officer’s email. The survey was distributed across diverse geographic locations within the United States across Arkansas, Massachusetts, New Mexico, Utah and Washington to capture a wide range of perspectives. Survey responses were collected between June 19, 2023 and September 26, 2023.

\subsubsection{Data Analysis}
We performed $\chi^2$ tests for each response against each of institution size, geographic location, and governance model (public or private). We also ran Pearson correlation tests for relation between need for policy, sentiment about AI tools, autonomy preference against administrators' experience length and student population. These tests were not significant.

\section{Results}

\begin{table}[h]
    \centering
    \begin{tabular}{|c|c|c|c|}
    \hline
        \textbf{State} & \textbf{High Schools} & \textbf{Higher Education} & \textbf{Total}  \\
        \hline
         Arkansas & 15 & 6 & 21\\
         Massachusetts & 13 & 3 & 16 \\
         New Mexico & 19 & 4 & 23 \\
         Utah & 18 & 5 & 23 \\
         Washington & 16 & 3 & 19 \\
         \hline
         Total & 81  & 21 & 102 \\  
        \hline
    \end{tabular}
    \caption{Survey responses by institution type and states}
    \label{tab:survey-result}
\end{table}

We received over 126 survey responses from across five states, some of which were partially completed. We had 102 complete surveys that we use for analysis for this study. Table~\ref{tab:survey-result} shows the number of responses from each state and type of educational institution.

\subsection{ RQ1: What is the current landscape of policies related to Generative AI in educational settings and what do these policies cover? }

The first research question investigates the presence and key components of policies or guidelines governing the use of emerging technologies such as Large Language Models (LLMs) and ChatGPT in educational environments.

\subsubsection{Existence of current policies}

\begin{figure}
    \centering
    \includegraphics[width=0.8\textwidth]{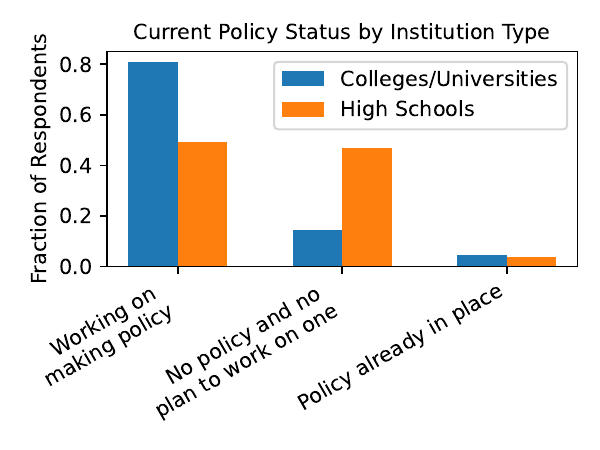}
    \caption{Current policy status by institution type}
    \label{fig:policy_status}
\end{figure}

A majority of respondents indicated either ongoing efforts to formulate generative AI-related policies or the existence of established policies. Specifically, over 80\% of higher education institutions reported active policy development, 5\% already have a policy, and 15\% have no plans to enact one. In contrast, only 50\% of high schools are in the process of policy formulation, while approximately 45\% neither have a policy nor plans to develop one. Figure~\ref{fig:policy_status} depicts these data. A statistically significant difference in policy status between high school and college was observed  $ \chi^2 (2, N = 102) = 0.7.44, p = .0.024$ indicating that high schools are less inclined to work on policies than higher educational institutions.

Having a very small sample size in each category doesn't allow us to analyze and understand differences between the categories, but we can still understand a lot  with the holistic review of the data.


 When asked if they need to have AI related policy, the prevailing sentiment among administrators was a critical need for these policies. Figure~\ref{fig:policy_necessary} shows the response on necessity of such policies. It can be seen that the necessity of AI related policy is almost universally agreed upon.
 

\subsubsection{Adequacy of Current Policies}

\begin{figure}[tbp]
    \centering
    \begin{subfigure}[b]{0.47\textwidth}
    \centering
    \includegraphics[width=\textwidth]{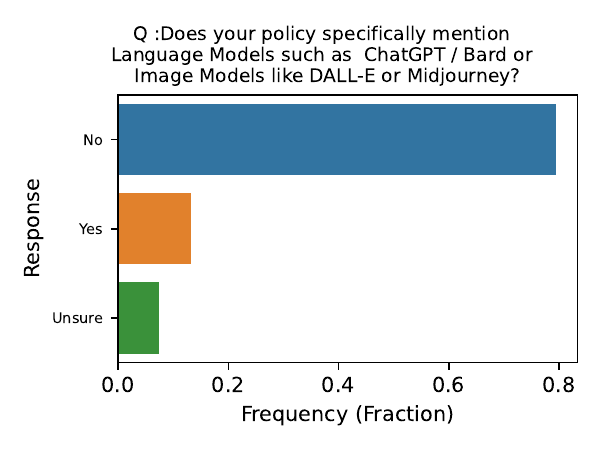}
    \caption{Specificity of in-place or in-progress policies in covering AI models}
    \label{fig:is-policy-specific}
    \end{subfigure}
    \hfill
    \begin{subfigure}[b]{0.47\textwidth}
        \centering
        \includegraphics[width=\textwidth]{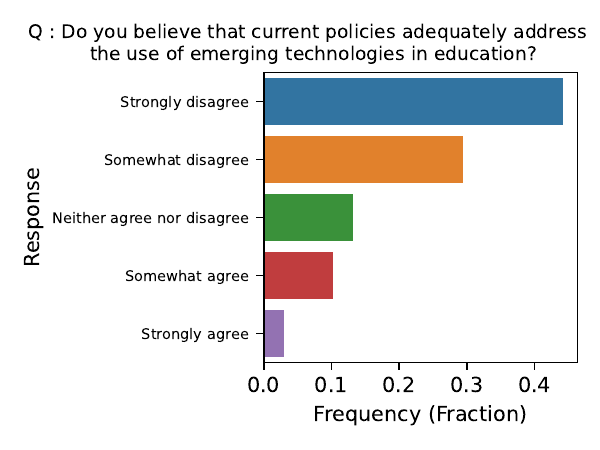}
        \caption{Adequacy of in-place or in-progress policies}
        \label{fig:is-policy-adequate}
     \end{subfigure}
        \caption{Administrators' responses on policy availability and adequacy}
        \label{fig:three graphs}
\end{figure}

Administrators who reported the existence or development of policies were subsequently asked about what is covered on their AI policies and their adequacy. The majority expressed that current or in-progress policies inadequately address the integration of emerging technologies. Figure~\ref{fig:is-policy-adequate} shows the administrators' perceptions on the adequacy of existing or in-development policies. Even for many policies currently in development, administrators think these policies are not adequate. 


Notably, only a small minority of these policies specifically mention LLMs like ChatGPT or Bard or image models like DALL-E. Figure~\ref{fig:is-policy-specific} illustrates these findings, suggesting an awareness gap in tailoring policies to specific technological challenges.

\begin{figure}[htbp]
    \centering
         \begin{subfigure}[b]{0.47\textwidth}
        \includegraphics[width=\textwidth]{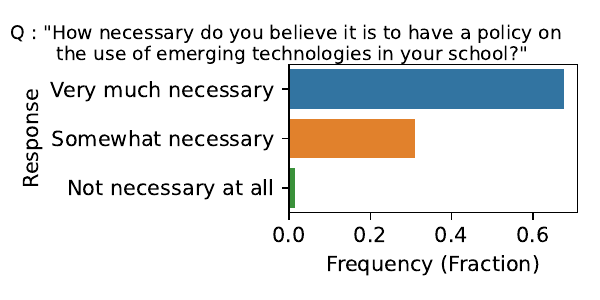}
        \caption{Necessity of AI policies}
        \label{fig:policy_necessary}
    \end{subfigure}
    \hfill
    \begin{subfigure}[b]{0.47\textwidth}
        \centering
        \includegraphics[width=\textwidth]{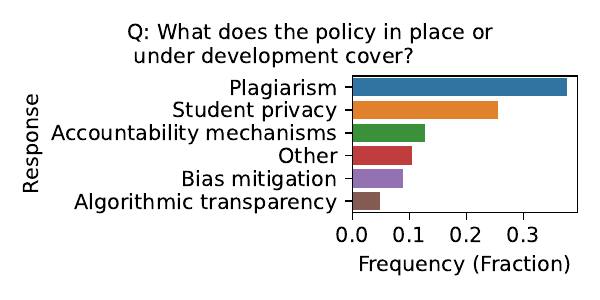}
        \caption{Components included in existing or in-development policies (multiple selection allowed)}
        \label{fig:policy_include}
    \end{subfigure}
        \caption{Administrators' responses on policy necessity and its components}
        \label{fig:three graphs}
\end{figure}



We also asked administrators what their current or in-progress policies covered. Existing policies most commonly address issues like plagiarism, while elements like bias mitigation and algorithmic transparency are less frequently covered. Ethical considerations' emerged as the most frequently cited motivation (25.6\%) for policy development or revision. This was followed by 'Ensuring student safety' (16.4\%). Least cited were 'Parental demand' and 'Teachers' demand', both under 5\%. Figure~\ref{fig:policy_include} indicates areas covered by current or in-progress policies. This indicates a perceived gap between existing governance mechanisms and the requirements for ethical and effective technology integration. Statistical tests revealed no significant associations between policy aspects and institution type, size, or location.


\subsection{RQ2 : What are the perceived needs for future policy formulation in relation to Generative AI, and what recommendations can be made for an effective ethical framework?}

Our second goal of this study was to understand the key elements that educational administrators believe should be included in a policy framework for the ethical use of emerging technologies like ChatGPT in education as well as their overall sentiment on the policy and gather any additional insight and recommendation from the administrators.



\subsubsection{Quantitative Analysis}
\begin{figure}
    \centering
         \begin{subfigure}[b]{0.47\textwidth}
         \centering
         \includegraphics[width=\textwidth]{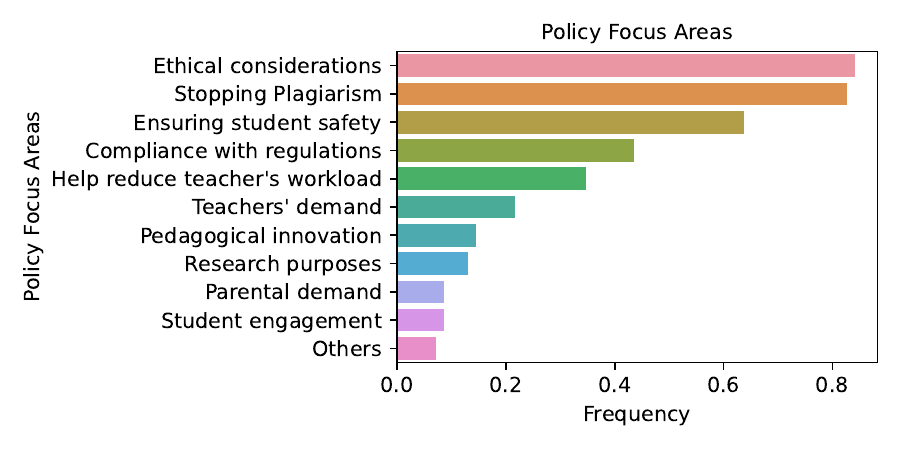}
        \caption{Focus area for policy identified by the administrators (Multiple selection allowed)}
        \label{fig:policy-focus}
    \end{subfigure}
    \hfill
    \begin{subfigure}[b]{0.47\textwidth}
        \includegraphics[width=\textwidth]{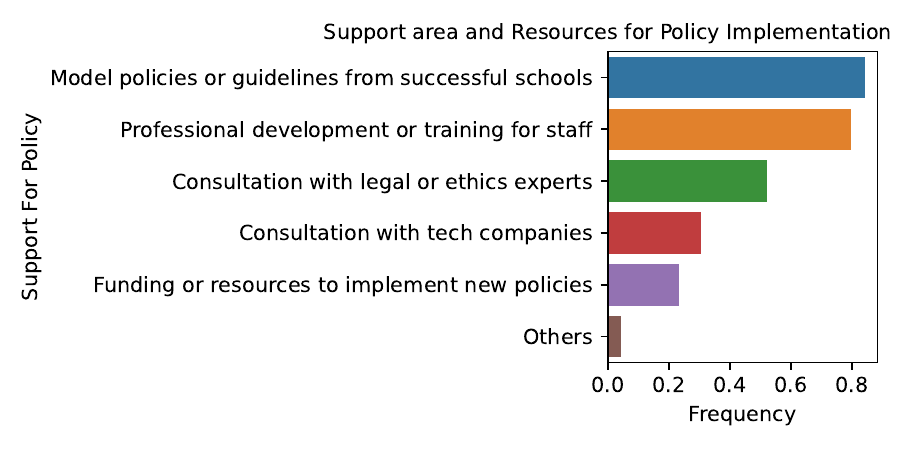}
        \caption{Important support resources identified by the administrators (Multiple selection allowed)}
        \label{fig:policy-support}
    \end{subfigure}
    \caption{Administrators' response in policy focus area and resources needed}
    \label{fig:three graphs}
    \end{figure}

Quantitatively, the focus was on the areas that respondents believe policies should primarily target and the kinds of support or resources they consider would be helpful for their institutions. Question "In which areas should policies for the use of emerging technologies in education primarily focus?" allowed multiple selections as well as free form text entry to capture administrators' focus area for policy making. Figure~\ref{fig:policy-focus} shows the policy focus area identified by school administers. The majority of respondents highlighted `Ethical Considerations' and `Stopping Plagiarism' as the top two areas, with over 80\% of responses, followed by ensuring students' safety and compliance with regulations.

We also asked the administrators about the support resources that would help them make or update generative AI related policies. The administers' answers are shown in Figure~\ref{fig:policy-support}. A model guidelines from successful school or district was the most commonly deemed useful resources, followed by professional development and staff training and legal/ethical consultations. Need for funding and resources and consultation with tech companies were also identified.

\subsubsection{Centralized Oversight vs. Decentralized Autonomy}
\begin{figure}
    \centering
         \begin{subfigure}[b]{0.47\textwidth}
         \centering
         \includegraphics[width=\textwidth]{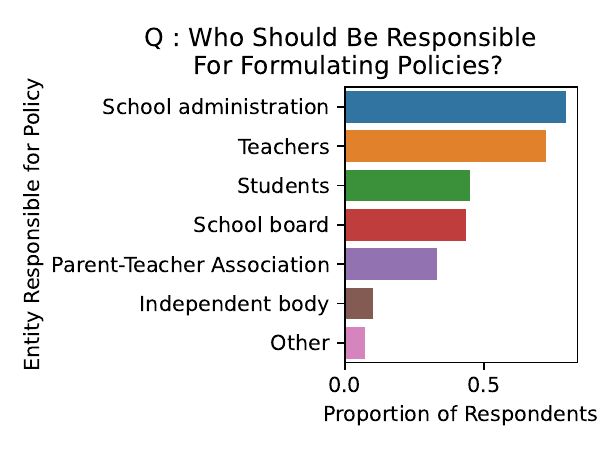}
    \caption{Responsible entity for policy-making (Multiple selection allowed) }
    \label{fig:responsible}
    \end{subfigure}
    \hfill
    \begin{subfigure}[b]{0.47\textwidth}
        \includegraphics[width=\textwidth]{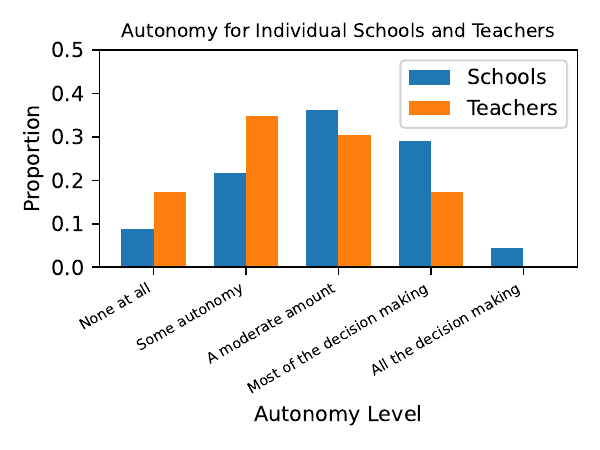}
        \caption{Autonomy (decision making) for individual school and individual teachers }
        \label{fig:autonomy}
    \end{subfigure}
    \caption{Administrators' responses on responsible entity and autonomy}
    \label{fig:three graphs}
    \end{figure}

The responses indicate a diverse perspective on who should be responsible and involved for formulating the policies governing the use of emerging technologies like ChatGPT in education. School administrators are seen as the most responsible entities, followed by teachers and students, along with school board and parent-teacher association. Figure~\ref{fig:responsible} shows the responsible entities identified for policy making purposes.


As for the autonomy and decision making given to schools and teachers, the respondents widely varied. For schools, the responses ranged from `none' to `all,' while the responses for teachers' decision-making ranged from `none' to `most.' Figure~\ref{fig:autonomy} shows the response for the question about autonomy and decision making power for schools and teachers respectively. Interestingly, none of the administrators responded that teacher should have all the decision making power (total autonomy).


Overall, the data suggests a preference for a collaborative approach to policy formulation and implementation that includes various stakeholders at different levels of governance.

\subsubsection{Qualitative Analysis}
The qualitative analysis was based on free-form text entries. While the number of responses was too limited to be able to perform a qualitative coding and analysis, they provided valuable insights. Respondents expressed concerns about the rapid advancements in technology and the need for policies to be flexible and adaptive, offering some explanation for why so few policies are currently in place. For example: 
 \begin{displayquote}
 \begin{tightitemize}
 \item  \textit{ "I believe that any policy should be reviewed and updated annually to keep up with advancements in technology."} \\
 \item  \textit{"The emerging AI platform will continue to grow and policies need to be flexible enough to adapt."} \\
 \item \textit{"This area of technology is moving so quickly that it's hard for policy to keep up."} 
      
 \end{tightitemize}
 \end{displayquote}

They also emphasized the importance of considering ethical implications, including potential biases in AI algorithms. One respondent noted, \textit{"I am concerned about the potential for bias in AI and think this should be addressed in any policy."} Others emphasized the ethical and privacy aspects, stating, \textit{"The policy must take into consideration the ethical implications of using AI in an educational setting.", and "I think privacy and data protection should be at the forefront of any policy concerning the use of AI technologies."} These quotes reflect the overarching sentiment that while technology is advancing rapidly, policies need to be robust yet flexible to adapt to these changes. Even when administrators are not clear what should be in the policy, they are quick to point out we have to be very careful on whatever policy we make:

 \begin{displayquote}
 \begin{tightitemize}

 \item  \textit{ "I'm not sure what the policy should contain, but I know it needs to be created carefully and with a lot of thought."} \\
 \item  \textit{"We are observing how AI impacts student learning and will be formulating a policy based on these findings. We are deliberately being very careful" }\\

 \end{tightitemize}
 \end{displayquote}

\subsubsection{Additional Observations}

\begin{figure}[htbp]
    \centering
         \begin{subfigure}[b]{0.48\textwidth}
         \centering
         \includegraphics[width=\textwidth]{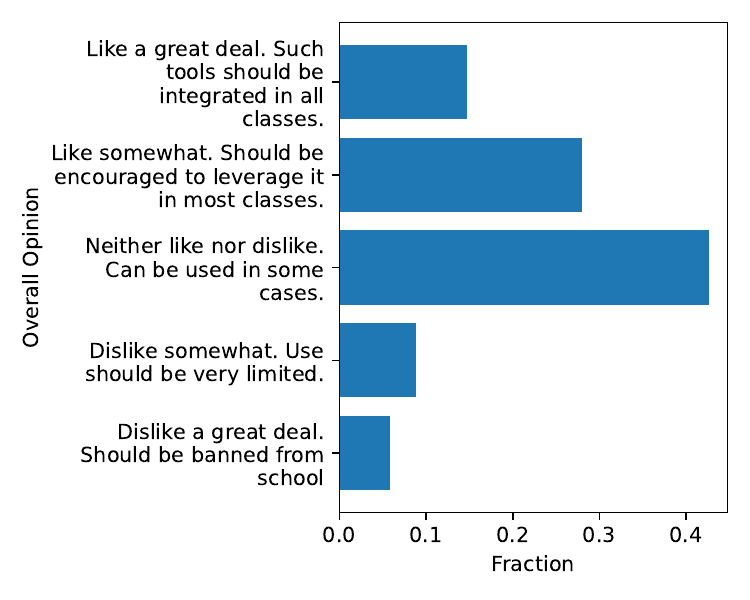}
        \caption{Overall opinion about AI in Education among school administrators }
        \label{fig:opinion}
    \end{subfigure}
    \hfill
    \begin{subfigure}[b]{0.48\textwidth}
        \includegraphics[width=\textwidth]{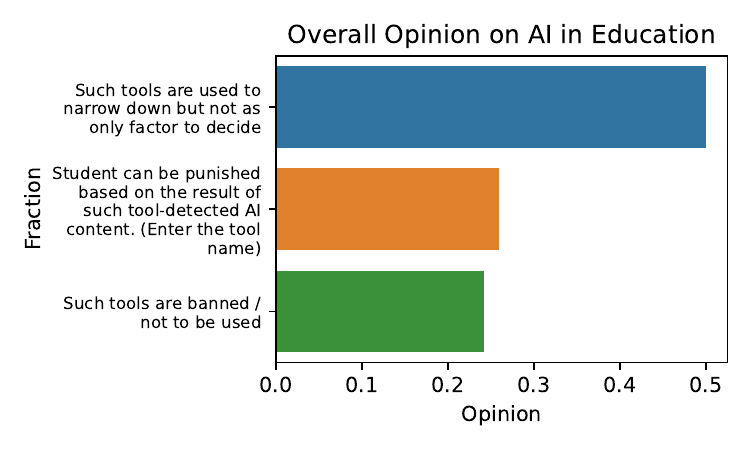}
        \caption{Overall opinion on existing AI generated content detection tool }
    \label{fig:detection-tool.pdf}
    \end{subfigure}
    \caption{Overall opinion on AI and AI detection tools}
    \label{fig:three graphs}
    \end{figure}

Additionally, we asked a couple of questions to understand the overall sentiment about AI tool as well as  sentiment about existing detection tools. Figure~\ref{fig:opinion} shows the overall opinion from these administrators. Most of the administrators are either indifferent or positive, and very few are not in favor of the technology. When asked about the use of existing tool that claim to detect AI-generated content, about half of the respondent were in favor of using such tools to narrow down, but not as a final arbiter of truth. The remaining respondents are almost evenly split between banning such tools and using such AI-detection tools. Figure~\ref{fig:detection-tool.pdf} shows the response for that question. We hypothesize that the high unreliability of these detection tools, their black-box nature and high cost of catching false positive are making the administrators take cautious approach towards detection tools.

\section{Conclusions and Discussion}

This study aimed to address two primary research questions (RQs) regarding the policy landscape for AI and LLM-based tools like ChatGPT in education. RQ1 explored the current state of policies and their coverage, revealing a significant push, especially in higher education, to develop guidelines. Yet, these policies often fall short of addressing the unique challenges of technologies like LLMs. We observed that high schools are less inclined to work on policies than higher educational institutions. The necessity of policy development was universally recognized among administrators, driven by ethical considerations and student safety, though areas like algorithmic transparency and bias mitigation were less emphasized, indicating gaps in existing frameworks.

RQ2 investigated the perceived needs for future policy formulation and proposed recommendations for an ethical framework. A preference for a collaborative, multi-stakeholder approach was evident, alongside the recognition that policies must be iterative and adaptable to keep pace with technological advances.

The findings indicate an active acknowledgment of AI and LLM's potential in education, alongside a nascent governance stage for their ethical and practical integration. Notably, the disparity in policy development between higher education and high schools—where about 40\% lack any policy efforts—points to potential resource or awareness discrepancies. This study underscores the critical gaps in policy adequacy and the necessity for policies to evolve alongside educational technologies. It emphasizes the importance of multi-stakeholder dialogues for creating governance mechanisms that are robust yet flexible enough to accommodate rapid technological changes.

The study concludes that the ethical and responsible integration of AI in education demands the continuous evolution of policies, practices, and attitudes. The findings of this study suggest for strategic, ethical, and collaborative governance, highlighting the imperative for developing comprehensive, adaptable policies to navigate the advancing landscape of AI technologies in educational settings.

\subsection{Future Work}

This study has laid important groundwork in understanding the state and direction of policies related to AI and LLMs in educational settings. However, several avenues for future research remain. The disparity in policy development between higher education and high schools warrants a more granular investigation. Future studies could focus on identifying the barriers and facilitators that influence policy-making at these disparate educational levels, possibly extending the research to include primary schools. Additionally, the evolving nature of AI and LLM technology itself calls for longitudinal studies that can track changes in administrative attitudes, policy adequacy, and implementation efficacy over time.

Another fruitful avenue for future work would be the exploration of multi-stakeholder perspectives, incorporating not just administrators but also teachers, students, and parents. Understanding these groups' attitudes and requirements could offer a more holistic view of what effective, comprehensive policies should entail. Investigations into the actual impact of AI and LLM-based tools on educational outcomes, based on these inclusive policies, could also provide valuable data for administrators and policy-makers.

\subsection{Threats to validity}
Our survey was not validated and no evaluation of reliability was made. Furthermore, all respondents were from institutions based in the United States, limiting external validity internationally. We did not collect any demographic information of participants. Finally, generative AI is a fast-moving technology and attitudes and policies are likely also changing quickly. This work represents a shapshot of policies and attitudes in mid-2023.

\bibliographystyle{splncs04}

\bibliography{refrences}

\begin{thebibliography}{10}
\providecommand{\url}[1]{\texttt{#1}}
\providecommand{\urlprefix}{URL }
\providecommand{\doi}[1]{https://doi.org/#1}

\bibitem{Adams2023Ethical}
Adams, C., Pente, P., Lemermeyer, G., Rockwell, G.: Ethical principles for artificial intelligence in k-12 education. Computers and Education: Artificial Intelligence  \textbf{4},  100131 (2023)

\bibitem{ai2019high}
AI, H.: High-level expert group on artificial intelligence (2019)

\bibitem{akgun2021artificial}
Akgun, S., Greenhow, C.: Artificial intelligence in education: Addressing ethical challenges in k-12 settings. AI and Ethics pp. 1--10 (2021)

\bibitem{ali2023constructing}
Ali, S., DiPaola, D., Williams, R., Ravi, P., Breazeal, C.: Constructing dreams using generative ai. arXiv preprint arXiv:2305.12013  (2023)

\bibitem{baidoo2023education}
Baidoo-Anu, D., Ansah, L.O.: Education in the era of generative artificial intelligence (ai): Understanding the potential benefits of chatgpt in promoting teaching and learning. Journal of AI  \textbf{7}(1),  52--62 (2023)

\bibitem{berendt2020ai}
Berendt, B., Littlejohn, A., Blakemore, M.: Ai in education: Learner choice and fundamental rights. Learning, Media and Technology  \textbf{45}(3),  312--324 (2020)

\bibitem{Chan2023Comprehensive}
Chan, C.K.Y.: A comprehensive ai policy education framework for university teaching and learning. International Journal of Educational Technology in Higher Education  \textbf{20}(1),  1--25 (2023)

\bibitem{chatila2018ethically}
Chatila, R., Firth-Butterfield, K., Havens, J.C.: Ethically aligned design: A vision for prioritizing human well-being with autonomous and intelligent systems version 2. University of southern California Los Angeles  (2018)

\bibitem{chen2022two}
Chen, X., Zou, D., Xie, H., Cheng, G., Liu, C.: Two decades of artificial intelligence in education. Educational Technology \& Society  \textbf{25}(1),  28--47 (2022)

\bibitem{chiu2023impact}
Chiu, T.K.: The impact of generative ai (genai) on practices, policies and research direction in education: a case of chatgpt and midjourney. Interactive Learning Environments pp. 1--17 (2023)

\bibitem{choi2023chatgpt}
Choi, J.H., Hickman, K.E., Monahan, A., Schwarcz, D.: Chatgpt goes to law school. Available at SSRN  (2023)

\bibitem{dwivedi2023so}
Dwivedi, Y.K., Kshetri, N., Hughes, L., Slade, E.L., Jeyaraj, A., Kar, A.K., Baabdullah, A.M., Koohang, A., Raghavan, V., Ahuja, M., et~al.: “so what if chatgpt wrote it?” multidisciplinary perspectives on opportunities, challenges and implications of generative conversational ai for research, practice and policy. International Journal of Information Management  \textbf{71},  102642 (2023)

\bibitem{filgueiras2023artificial}
Filgueiras, F.: Artificial intelligence and education governance. Education, Citizenship and Social Justice p. 17461979231160674 (2023)

\bibitem{flogie2023artificial}
Flogie, A., Krabonja, M.V.: Artificial intelligence in education: developing competencies and supporting teachers in implementing ai in school learning environments. In: 2023 12th Mediterranean Conference on Embedded Computing (MECO). pp.~1--6. IEEE (2023)

\bibitem{garshi2020smart}
Garshi, A., Jakobsen, M.W., Nyborg-Christensen, J., Ostnes, D., Ovchinnikova, M.: Smart technology in the classroom: a systematic review. prospects for algorithmic accountability. arXiv preprint arXiv:2007.06374  (2020)

\bibitem{Gillani2023Unpacking}
Gillani, N., Eynon, R., Chiabaut, C., Finkel, K.: Unpacking the “black box” of ai in education. Educational Technology \& Society  \textbf{26}(1),  99--111 (2023)

\bibitem{halaweh2023chatgpt}
Halaweh, M.: Chatgpt in education: Strategies for responsible implementation  (2023)

\bibitem{holmes2021ethics}
Holmes, W., Porayska-Pomsta, K., Holstein, K., Sutherland, E., Baker, T., Shum, S.B., Santos, O.C., Rodrigo, M.T., Cukurova, M., Bittencourt, I.I., et~al.: Ethics of ai in education: Towards a community-wide framework. International Journal of Artificial Intelligence in Education pp. 1--23 (2021)

\bibitem{Kooli2023Chatbots}
Kooli, C.: Chatbots in education and research: A critical examination of ethical implications and solutions. Sustainability  \textbf{15}(7), ~5614 (2023)

\bibitem{li2023risk}
Li, S., Gu, X.: A risk framework for human-centered artificial intelligence in education. Educational Technology \& Society  \textbf{26}(1),  187--202 (2023)

\bibitem{lo2023impact}
Lo, C.K.: What is the impact of chatgpt on education? a rapid review of the literature. Education Sciences  \textbf{13}(4), ~410 (2023)

\bibitem{memarian2023fairness}
Memarian, B., Doleck, T.: Fairness, accountability, transparency, and ethics (fate) in artificial intelligence (ai), and higher education: A systematic review. Computers and Education: Artificial Intelligence p. 100152 (2023)

\bibitem{miao2021ai}
Miao, F., Holmes, W., Huang, R., Zhang, H., et~al.: AI and education: A guidance for policymakers. UNESCO Publishing (2021)

\bibitem{Nguyen2023Ethical}
Nguyen, A., Ngo, H.N., Hong, Y., Dang, B., Nguyen, B.P.T.: Ethical principles for artificial intelligence in education. Education and Information Technologies  \textbf{28}(4),  4221--4241 (2023)

\bibitem{nigam2021systematic}
Nigam, A., Pasricha, R., Singh, T., Churi, P.: A systematic review on ai-based proctoring systems: Past, present and future. Education and Information Technologies  \textbf{26}(5),  6421--6445 (2021)

\bibitem{ouyang2021artificial}
Ouyang, F., Jiao, P.: Artificial intelligence in education: The three paradigms. Computers and Education: Artificial Intelligence  \textbf{2},  100020 (2021)

\bibitem{pradana2023discussing}
Pradana, M., Elisa, H.P., Syarifuddin, S.: Discussing chatgpt in education: A literature review and bibliometric analysis. Cogent Education  \textbf{10}(2),  2243134 (2023)

\bibitem{sahlgren2023politics}
Sahlgren, O.: The politics and reciprocal (re) configuration of accountability and fairness in data-driven education. Learning, Media and Technology  \textbf{48}(1),  95--108 (2023)

\bibitem{sattelmaier2023towards}
Sattelmaier, L., Pawlowski, J.M.: Towards a generative artificial intelligence competence framework for schools. In: Proceedings of the International Conference on Enterprise and Industrial Systems (ICOEINS 2023). vol.~270, p.~291. Springer Nature (2023)

\bibitem{cosnCoSNIssues2019}
for School~Networking, C.: Cosn strategic plan 2019-2022 (2019), [Accessed 06-10-2023]

\bibitem{cosnCoSNIssues2023}
for School~Networking, C.: {C}o{S}{N} {I}ssues {G}uidance on {A}{I} in the {C}lassroom | {C}o{S}{N} --- cosn.org (2023), [Accessed 06-10-2023]

\bibitem{sullivan2023chatgpt}
Sullivan, M., Kelly, A., McLaughlan, P.: Chatgpt in higher education: Considerations for academic integrity and student learning  (2023)

\bibitem{Uunona2023Leveraging}
Uunona, G.N., Goosen, L.: Leveraging ethical standards in artificial intelligence technologies: A guideline for responsible teaching and learning applications. In: Handbook of Research on Instructional Technologies in Health Education and Allied Disciplines, pp. 310--330. IGI Global (2023)

\bibitem{whalen2023chatgpt}
Whalen, J., Mouza, C., et~al.: Chatgpt: Challenges, opportunities, and implications for teacher education. Contemporary Issues in Technology and Teacher Education  \textbf{23}(1),  1--23 (2023)

\bibitem{Zhai2021Review}
Zhai, X., Chu, X., Chai, C.S., Jong, M.S.Y., Istenic, A., Spector, M., Liu, J.B., Yuan, J., Li, Y.: A review of artificial intelligence (ai) in education from 2010 to 2020. Complexity  \textbf{2021},  1--18 (2021)

\end{thebibliography}

\end{document}